\documentstyle[multicol,epsf,aps]{revtex}
\begin{document}
\draft
\title{Supernova electron capture rates on odd-odd nuclei}
\author{K. Langanke and G. Mart\'{\i}nez-Pinedo}  

\address{Institute for Physics and Astronomy, University of {\AA}rhus,
  Denmark and Theoretical Center for Astrophysics, University of
  {\AA}rhus, Denmark}

\date{\today}

\maketitle

\begin{abstract}
  At densities between $10^8$ and $10^{10}$ g/cm$^3$ electron capture
  in a presupernova collapse is believed to mainly occur on odd-odd
  nuclei.  We have derived the rates for six of the most important
  electron capturing nuclei, $^{54,56,58}$Mn and $^{56,58,60}$Co,
  based on calculations of the Gamow-Teller strength distributions for
  the ground states and first excited states.  These calculations have
  been performed by shell model diagonalization in the $pf$ shell
  using a recently modified version of the KB3 interaction.  The shell
  model rates are noticeably smaller than the presently adopted rates
  as the latter have been derived by placing the Gamow-Teller (GT)
  resonance at too low excitation energies.
\end{abstract}

\pacs{PACS numbers: 26.50.+x, 23.40.-s, 21.60Cs, 21.60Ka}

\begin{multicols}{2}

If the core of a massive star exceeds the appropriate Chandrasekhar
mass, electron degeneracy pressure cannot longer stabilize the center
and it collapses. In this early stage of the collapse electron capture
plays an essential role. At first, it reduces the number of leptons
per baryons $Y_e$ and hence the pressure which the electron gas can
stem against the collapse. Secondly, the densities are still low
enough for the neutrinos, produced by the electron capture process, to
leave the star and thus to carry some energy away and cool the core.
Thus, both effects conspire to accelerate the collapse.  The
importance of electron capture for the presupernova collapse is for
example discussed in \cite{Bethe78,Bethe90}.

Core collapse models employ the electron capture rates by Fuller,
Fowler and Newman (FFN) \cite{FFN} who have systematically estimated
the rates for nuclei in the mass range $A=45-60$. The FFN rates are
derived from two distinct contributions. At first the authors
estimated the Gamow-Teller (GT) contribution to the rate by a
parametrization on the basis of the independent particle model. The
rate estimate has then be completed by an empirical contribution
placed at zero excitation energy simulating low-lying transitions.
After experimental evidence suggested that the GT strength is quenched
with respect to the independent particle model, the FFN rates have
been updated by Aufderheide {\it et al.} \cite{Aufderheide} by
quenching of the Gamow-Teller strength by an overall factor of two.
Furthermore these authors simulated the low-lying transitions by the
same $ft$-value for all nuclei, while FFN adopted specific values for
individual nuclei.

Using their own rate estimates, Aufderheide {\it et al.} have ranked
the most important electron capturing nuclei -- defined by the product
of the abundance of a given nucleus and its electron capture rate --
along a stellar trajectory for core collapse densities
$\rho=10^{7}-10^{10}$ g/cm$^3$ \cite{Aufderheide}.  They find that for
densities $\rho_7 > 10$ ($\rho_7$ measures the density in $10^7$
g/cm$^3$) electrons are most effectively captured by odd-odd nuclei.
In particular, with increasing density, $^{54}$Mn, $^{60}$Co and
$^{58}$Mn are subsequently the top-ranked nuclei, which decrease $Y_e$
by electron capture most effectively.  It is important to note that
the FFN rates agree with those of Ref.  \cite{Aufderheide} within a
factor of two, thus also showing the dominance of electron capture by
odd-odd nuclei in this stage of the collapse. It is the aim of this
paper to show that this finding results from a misplacement of the
Gamow-Teller resonance position in the parametrizations used by FFN
\cite{FFN} and Aufderheide {\it et al.}  \cite{Aufderheide}.

Unfortunately there exists no experimental information about the
Gamow-Teller strength distribution for odd-odd nuclei in the
$pf$-shell.  Therefore our conclusions have to be entirely based on
theory.  As our theoretical model of choice we adopt the interacting
shell model.  Recent progress allows now for virtually converged
calculations of the Gamow-Teller strength for all nuclei in the $pf$
shell \cite{Nowacki}.  In fact, it has been shown that the shell model
studies reproduce all measured GT distributions for nuclei in the mass
range $A=50-64$, which is important for the core collapse phase we are
concerned with here \cite{Martinez99} (also see \cite{Radha97}).  The
nuclei, for which GT data exist, comprise both even-even ones (e.g.
$^{54-58}$Fe, $^{58-64}$Ni) and odd-A nuclei ($^{51}$V, $^{55}$Mn,
$^{59}$Co).  For the following discussion it is important to note that
the calculations \cite{Radha97,Martinez99}, in concordance with data
\cite{gtdata1,gtdata2,gtdata3,gtdata4,gtdata5}, showed systematic
misplacements of the GT resonance strength in the parametrizations
used by FFN \cite{FFN} and subsequently by Aufderheide {\it et al.}
\cite{Aufderheide}.  These authors placed the centroid of the GT
strength at too low excitation energies in the daughter nucleus for
electron capture on odd-A nuclei, while they assumed too high
excitation energies for capture on even-even nuclei.

Motivated by the successful application to even-even and odd-A nuclei
\cite{Radha97,Martinez99,Martinez98} we assume that the interacting
shell model will also describe the GT distribution for odd-odd nuclei
well.  Thus we have calculated the GT strength distribution for the
six odd-odd nuclei $^{54,56,58}$Mn and $^{56,58,60}$Co on the basis of
a shell model diagonalization approach in the $pf$ shell. As residual
interaction we adopted the recently modified version of the KB3
interaction which corrects the slight inefficiencies in the KB3
interaction around the $N=28$ subshell closure \cite{Ni56}.  In fact
the modified KB3 interaction i) reproduces all measured GT strength
distributions very well and ii) describes the experimental level
spectrum of the nuclei studied here quite accurately
\cite{Martinez99,Ni56}.  Due to the very large m-scheme dimensions
involved, the GT strength distributions have been calculated in
truncated model spaces which fulfills the Ikeda sum rule and allowed a
maximum of 4 particles from the lowest independent particle model
configuration to be excited from the $f_{7/2}$ shell to the rest of
the $pf$ shell in the final nucleus. At this level of truncation the GT
strength distribution is virtually converged and the total GT strength
agrees with the exact value typically within $10\%$.

As $0\hbar\omega$ shell model calculations, i.e.  calculations
performed in one major shell, overestimate the experimental GT
strength by a universal factor
\cite{Wildenthal,Langanke95,Martinez96}, we have scaled our GT
strength distribution by this factor, $(0.74)^2$.

We have performed 33 Lanczos iterations for each final angular
momentum, which are usually sufficient to converge in the states at
excitation energies below $E= 3$ MeV.  At higher excitation energies,
$E>3$ MeV, the calculated GT strengths represent centroids of
strengths, which in reality are splitted over many states.  For
calculating the electron capture rate, however, a resolution of this
strength at higher energies is unimportant.

Once the GT distributions are known the electron capture rate can be
calculated as outlined in \cite{FFN,Aufderheide}. We note, however,
that in the core collapse environment the capture process occurs at
finite temperature ($T \approx (4-7) \cdot 10^9$ K at the densities we
are concerned with here \cite{Aufderheide}). Thus we have included in
our rate calculations also the capture from thermally excited states
in the parent nucleus at excitation energies below 1 MeV. As the
ground state spin of the even-even daughter nuclei is $J=0$ (which is
often strongly mismatched with the spins of the low-lying states in
the odd-odd parent), we have included for all parent nuclei capture
from at least one $1^+$ state.  In $^{56}$Co the lowest excited $J=1$
state is at 1.71 MeV.  For the excitation energies we have used the
experimental values rather than the shell model results, although they
usually agree within 100 keV.  Further, if the energy of a specific
final state is known experimentally we have used this value. For the
mass splittings between daughter and parent nucleus we adopt the
experimental values.

In Fig. 1 we have plotted the GT strength distributions for the ground
states of the odd-odd nuclei $^{54,56,58}$Mn and $^{56,58,60}$Co. As
the ground states of these nuclei have spin $J \ne 0$, GT transitions
can lead to final states with angular momentum $J-1,J$ and $J+1$; the
figure shows the three individual strength distributions. In Fig. 2 we
compare the ground state GT distribution with those of the excited
states, adopting $^{54}$Mn as a typical example.  Several observations
can be derived from the two figures.

As the most striking feature we find that the shell model places the
centroid of the GT distribution at higher energies than adopted in the
parametrizations of FFN \cite{FFN} and Aufderheide {\it et al.}
\cite{Aufderheide}.  To be more quantitative we have calculated the GT
centroids $E_{GT}$ for the various ground states, averaged over the
three possible final states, and find $E_{GT}=7.15$ MeV ($^{54}$Mn),
5.9 MeV ($^{56}$Mn), 5.5 MeV ($^{58}$Mn), 8.2 MeV ($^{56}$Co), 7.35
MeV ($^{58}$Co) and 6.35 MeV ($^{60}$Co). These values are typically
more than 2 MeV higher than the parametrizations used in
\cite{FFN,Aufderheide} (see Fig. 1). Only for $^{56}$Mn the difference
is only about 0.6 MeV.  For the total $B(GT)$ values (in units of
$g_A^2$, where $g_A$ is the axialvector coupling constant) we
calculate 4.4 (8.6), 2.7 (8.6), 1.5 (7.2) for $^{54,56,58}$Mn and 7.7
(12.0), 5.9 (12.0), 3.7 (10.0) for $^{56,58,60}$Co, where the numbers
in parentheses are the independent particle values. Considering the
additional reduction of the total $B(GT)$ strength related to the
universal renormalization factor $(0.74)^2$ we conclude that the GT
strength is stronger quenched than even assumed in \cite{Aufderheide}.
As a consequence of the differences in the total strength and in the
position of the centroid one expects that the bulk of the GT
transition will contribute less to the electron capture rates than
assumed previously.

The shell model gives very weak transition strengths to low-lying
states in the daughter nucleus. In particular, the rather large ground
state spins of the odd-odd nuclei (except for $^{58}$Mn) allow only
transitions to excited states in the daughter; this is particularly
drastic for $^{56}$Co ($J=4$) and $^{60}$Co ($J=5$). Viceversa,
transitions to the daughter ground state are only possible from $J=1$
states in the parent which are usually (except for $^{58}$Mn)
suppressed by the thermal Boltzmann factor.

In previous compilations \cite{FFN,Aufderheide} the electron capture
rate at finite temperature has been calculated employing the so-called
Brink hypothesis. This assumes that the GT strength distribution on
excited states is the same as for the ground state, only shifted by
the excitation energy of the state \cite{Aufderheide91}.  As can be
seen in Fig. 2, this assumption is valid for the bulk of the GT
strength, but is not applicable for the individual transitions to
states at low-excitation energy in the daughter. To be more
quantitative, we have calculated the GT centroids ($E_{GT}$) for
various states in $^{56}$Co: the ground state, the $J=3$ state at 0.22
MeV (0.16 MeV), the $J=5$ state at 0.60 MeV (0.57 MeV), the $J=2$
state at 1.04 MeV (0.97 MeV) and the $J=1$ state at 1.94 MeV (1.72
MeV) where we have compared our shell model excitation energies $E_x$
with the experimental values given in parentheses. A measure for the
validity of the Brink hypothesis is then given by the difference
$E_{GT}-E_x$ and we find within 60 keV the same values for this
quantity for the lowest $J=2$-5 states ($\approx 8.2$~MeV); the
difference for the $J=1$ state is smaller (7.4 MeV) mainly caused by
the strong transition to the ground state of $^{56}$Mn which exhaust
about 25$\%$ of the total $B(GT)$ strength found in the transition to
final $J=0$ states.  The applicability of Brink's hypothesis has
already been discussed in \cite{Aufderheide93a,Martinez98}.

The shell model results for the low-lying strength indicate that the
empirical $ft$ value adopted in~\cite{Aufderheide} ($B(GT)=0.1$) to
simulate low-lying transitions is too large.

The calculated electron capture rates for the six nuclei are shown in
Fig.~3 as function of temperature ($T_9 $ measures the temperature in
$10^9$ K) and at those densities at which the individual nuclei have
been identified in Ref. \cite{Aufderheide} as most important for the
electron capture process.  For the chemical potential we use the
approximation \cite{Aufderheide90}

\begin{equation}
\mu_e = 1.11 (\rho_7 Y_e)^{1/3}\left[1+\left(\frac{\pi}{1.11}\right)^2
\frac{T^2}{\left(\rho_7 Y_e\right)^{2/3}}\right]^{-1/3}\;.
\end{equation}

The present shell model rates are compared to the FFN rates at the
same densities. Additionally the figure indicates the rate of Ref.
\cite{Aufderheide} taken from their Tables 15-17.

As expected from the discussion above, the shell model rates are
significantly smaller than the rates of Ref. \cite{Aufderheide} and
typically also than the FFN rates. The only exception here is the
capture rate on $^{58}$Mn, where the FFN rates are smaller than ours
at low temperatures. This, however, is due to the fact that the mass
of $^{58}$Cr, the daughter of $^{58}$Mn, has not been known
experimentally at the time the FFN rates have been derived and these
authors used the mass systematics from Seeger and Howard
\cite{Seeger}.  As discussed in \cite{Aufderheide}, this resulted in a
quite different $Q$-value for this reaction.  We note that it is
particularly interesting to compare the rates in Fig.  3 at the
density and temperature combinations quoted for the results from Ref.
\cite{Aufderheide} as they are along the stellar trajectory at which
the collapse is expected to occur if the FFN rates are employed.  For
the Mn isotopes the shell model rates are smaller by factors 4
($^{56}$Mn) to 12 ($^{54}$Mn). The rather small ratio for $^{56}$Mn
reflects the fact that this is the case among the nuclei studied here
where the FFN and shell model centroids of the Gamow-Teller
distributions agree best.  In passing we note that for $^{56}$Mn the
low-lying strength assumed in Ref. \cite{Aufderheide} strongly exceeds
the shell model value and the one estimated by FFN. For the Co
isotopes the reduction of the rates compared to FFN is drastic ranging
from a factor 30 ($^{58}$Co) to 400 ($^{60}$Co), mainly caused by the
misplacement of the GT centroid in the previous parametrizations.  The
shell model calculations certainly do not substantiate the large
amount of the capture rate attributed to the GT resonance in Ref.
\cite{Aufderheide}.  The reductions for $^{54}$Mn and $^{60}$Co are
quite relevant as Aufderheide {et al.} state that these nuclei
contribute about $20\%$ and $50\%$ to the change of $Y_e$ at certain
stages of the collapse \cite{Aufderheide}.

We do believe that the odd-odd nuclei studied here reflect a typical,
rather than an exceptional sample. Accepting this point of view one is
lead to the conclusion that the current compilations of electron
capture rates are based on a parametrization which places the GT
centroid for odd-odd parent nuclei at too low excitation energies.
Consequently the electron capture rates on odd-odd nuclei, as
recommended in \cite{FFN} and \cite{Aufderheide}, are too large.
Judging the overestimation of the rates from the six nuclei studied
here, a reduction of the rates by about an order of magnitude might be
anticipated.  Previous shell model studies indicate that the
recommended capture rates for odd-A nuclei are also likely too large
due to a similar misplacing of the GT centroid, while the FFN rates
are roughly confirmed for capture on even-even nuclei
\cite{Martinez98,Dean97}. Summarizing these indications, one expects
that the total electron capture rate relevant for the presupernova
collapse at densities $\rho_7 \le 1000$ is smaller than currently
believed. As a consequence of a slower electron capture rate, the core
radiates less energy away by neutrino emission, keeping the core on a
trajectory with higher temperature and entropy. However, drawing
conclusions about possible effects which lower electron capture rates
might have on the collapse mechanism, in particular on the size of the
homologous core, are premature and prohibited at this stage. First,
one has to compile a complete set of shell model based capture rates
for all relevant nuclei. Secondly, during the collapse electron
capture has to compete with $\beta$-decay and preliminary results
indicate that the shell model roughly confirms the total FFN rates
\cite{Martinez98b}.  If true, electron capture and $\beta$ decay rates
balance during the stellar collapse and might lead to a cooling of the
star without changing its $Y_e$ value.  This possibility has already
been suggested in Ref. \cite{URCA} on the basis of a few experimental
GT distributions.

In summary, we have performed state-of-the-art large-scale shell model
diagonalization calculations to determine the presupernova electron
capture rates on selected odd-odd nuclei $^{54,56,58}$Mn and
$^{56,58,60}$Co, which are believed to be the most important
``electron poisons'' during the stellar collapse.  Our calculations
suggest that the previous compilations of these rates placed the GT
centroids for odd-odd parent nuclei at too low excitation energies in
the daughter. As a consequence we calculate significantly smaller
electron capture rates for all studied odd-odd nuclei than given in
the standard compilation of Fuller, Fowler and Newman \cite{FFN}.

\acknowledgements

Discussion with W.C. Haxton are gratefully acknowledged.  This work
was supported in part by the Danish Research Council.  Computational
resources were provided by the Center for Advanced Computational
Research at Caltech.

\end{multicols}

\begin{multicols}{2}
\narrowtext

\begin{figure}
  \begin{center}
    \leavevmode
  \epsfxsize=0.45\textwidth
  \epsffile{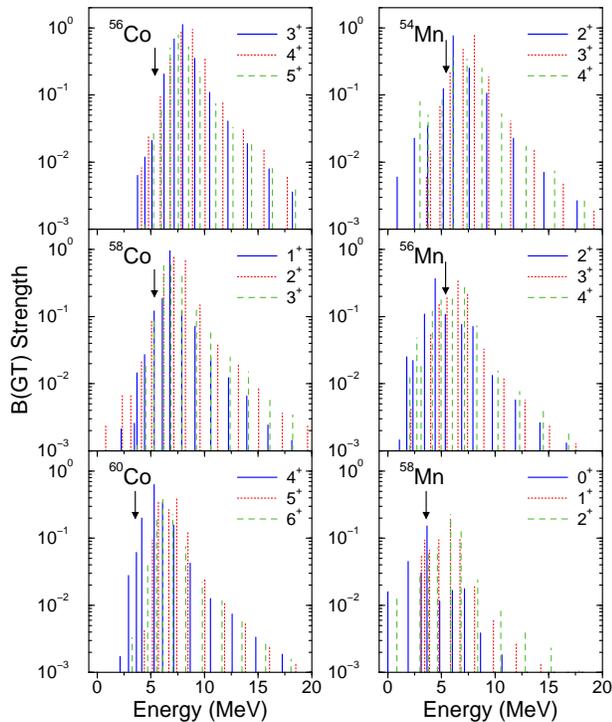}
  \caption{Gamow-Teller strength distributions for the ground states of
    $^{54,56,58}$Mn and $^{56,58,60}$Co, as calculated in the present
    shell model approach.  The distributions for the various final
    angular momenta are given separately. The arrows indicate the
    energies at which the compilations of Fuller, Fowler and Newman
    \protect\cite{FFN} placed the centroid of the GT strength.  The
    energy scale refers to excitation energies in the daughter
    nucleus.}
  \label{fig1}
\end{center}
\end{figure}

\begin{figure}
  \begin{center}
    \leavevmode
  \epsfxsize=0.45\textwidth
  \epsffile{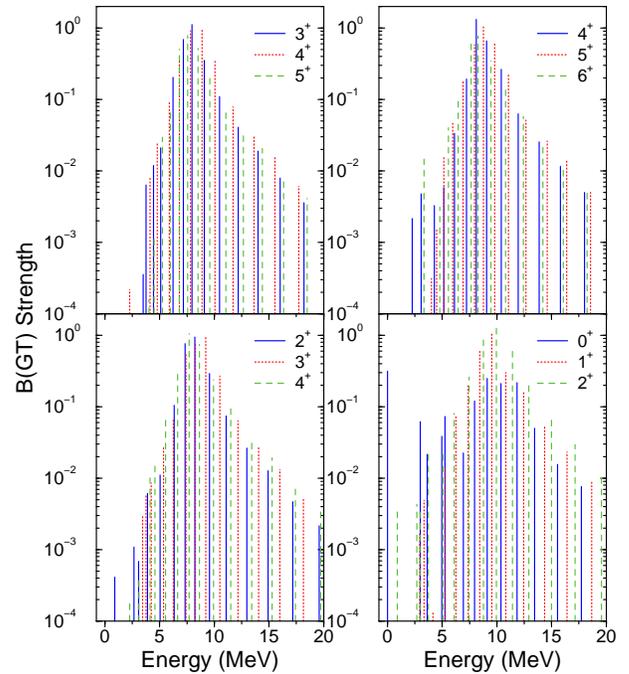}
  \caption{Gamow-Teller strength distribution for the $^{56}$Co ground
    state ($J=4$) and the first excited states with angular momentum
    $J=3,5$ and $J=1$. Experimentally these states are found at an
    excitation energy of 0.158 MeV, 0.576 MeV, and 1.72 MeV,
    respectively.}
  \label{fig2}
\end{center}
\end{figure}

\end{multicols}

\begin{figure}
  \begin{center}
    \leavevmode
    \epsfxsize=0.45\textwidth
    \epsffile{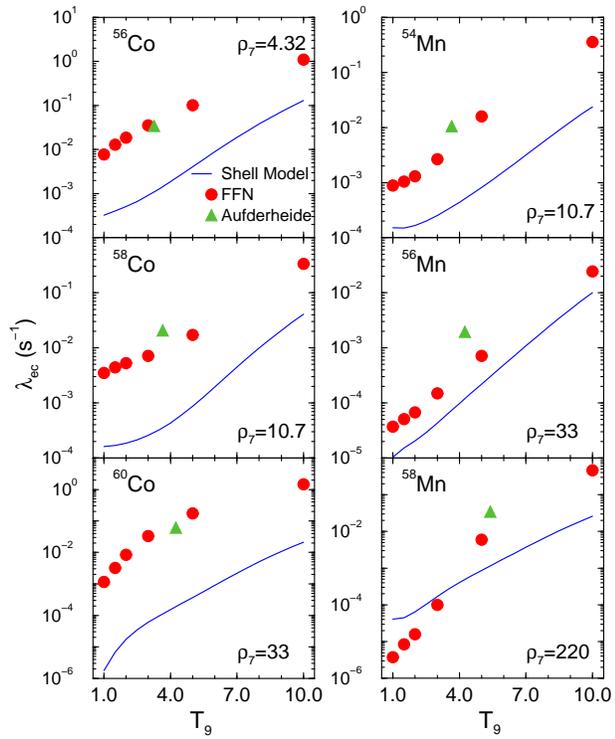}
    \caption{Electron capture rates on $^{54,56,58}$Mn and
    $^{56,58,60}$Co as a function of temperature and at selected
    densities at which these nuclei are most important for electron
    capture in the presupernova core collapse as suggested by Ref.
    \protect\cite{Aufderheide}.  The solid line shows the present
    shell model results, the dots give the FFN rates
    \protect\cite{FFN}, while the triangles are rates taken from
    Tables 15-17 in \protect\cite{Aufderheide}.}
  \label{fig3}
  \end{center}
\end{figure}

\end{document}